\documentclass[aps,twocolumn,superscriptaddress]{revtex4-2}
\usepackage{amsmath,amssymb,bm}
\usepackage{graphicx}
\usepackage{xcolor}

\begin{document}

\title{Electrically driven Rabi dynamics of magnetic-field-induced corner states in a two-dimensional topological insulator} 

\author{D.~V. Khomitsky}
\email{khomitsky@phys.unn.ru}
\affiliation{National Research Lobachevsky State University of Nizhny Novgorod, Department of Physics, 23 Gagarin Avenue, 603022 Nizhny Novgorod, Russia}
\affiliation{National Research University Higher School of Economics, Department of Informatics, Mathematics and Computer Sciences, 136 Rodionov Street, 603093 Nizhny Novgorod, Russia}

\author{E.~A. Lavrukhina}
\affiliation{National Research Lobachevsky State University of Nizhny Novgorod, Research Physicotechnical Institute, 23 Gagarin Avenue, 603022 Nizhny Novgorod, Russia}

\author{D.~P. Krasavin}
\affiliation{National Research Lobachevsky State University of Nizhny Novgorod, Department of Physics, 23 Gagarin Avenue, 603022 Nizhny Novgorod, Russia}

\author{S. S.~Krishtopenko}
\affiliation{Laboratoire Charles Coulomb (L2C), UMR 5221 CNRS-Universit\'{e} de Montpellier, F-34095 Montpellier, France}
\affiliation{Julius-Maximilians-Universit\"{a}t W\"{u}rzburg, Physikalisches Institut and W\"{u}rzburg-Dresden Cluster of Excellence ctd.qmat, Lehrstuhl f\"{u}r Technische Physik, Am Hubland, 97074 W\"{u}rzburg, Deutschland}

\author{F.~Teppe}
\email{frederic.teppe@umontpellier.fr}
\affiliation{Laboratoire Charles Coulomb (L2C), UMR 5221 CNRS-Universit\'{e} de Montpellier, F-34095 Montpellier, France}

\date{\today}

\begin{abstract}
We study coherent electric manipulation of magnetic-field-induced localized states at a double kink of a helical edge in a HgTe/CdHgTe quantum well. An in-plane magnetic field opens a gap in the one-dimensional edge spectrum, while changes in the edge orientation generate localized in-gap states at the kinks. We show that, for suitable geometry and magnetic-field direction, two such states form an effective lithographically defined two-level subsystem. Using an edge-state model that includes both the localized levels and the fully propagating 1D edge states outside the magnetic-field-induced gap, we calculate the electric-dipole matrix elements and solve the time-dependent problem under resonant driving. The resulting dynamics exhibits Rabi oscillations with linear frequencies of $20$--$40$~GHz for realistic parameters. We find that the propagating 1D edge states provide a leakage channel whose strength is strongly controlled by the driving amplitude: reducing the electric field suppresses leakage below the percent level while preserving GHz-scale coherent oscillations. These results establish a route from magnetic-field-induced corner-state physics to electrically driven two-level dynamics in a realistic two-dimensional topological-insulator edge geometry.
\end{abstract}

\keywords{topological insulators, HgTe/CdHgTe quantum well, corner states, Rabi oscillations}
\maketitle

\section{Introduction}
Two-dimensional (2D) topological insulators (TIs) realizing the quantum spin Hall phase are characterized by an insulating bulk gap and gapless one-dimensional (1D) helical edge states~\cite{TI1,TI2,Bernevig2006,Konig2007,Roth2009}. Their key transport signature is a quantized conductance at zero magnetic field, associated with current carried by counterpropagating spin-momentum-locked edge modes. HgTe/CdHgTe quantum wells (QWs) provided the first semiconductor platform in which the quantum spin Hall phase was both theoretically predicted~\cite{Bernevig2006} and subsequently observed experimentally at millikelvin temperatures~\cite{Konig2007,Roth2009}. Since then, the family of experimentally realized 2D TIs has expanded beyond HgTe-based systems. Quantized edge transport has been reported in monolayer WTe$_2$ up to temperatures of order $100$~K~\cite{Wu2018}, and, more recently, in inverted three-layer InAs/GaInSb/InAs QWs up to $60$~K~\cite{KrishtopenkoTeppe2018SciAdv,Meyer2025SciAdv}. These developments establish semiconductor QWs and related van der Waals systems as versatile platforms for studying and controlling helical edge states.

The development of boundary-state concepts in topological phases has led to the notion of higher-order topological insulators (HOTIs)~\cite{Benalcazar2017,Benalcazar2017a,Langbehn2017,Schindler2018,Schindler2018a,Xie2021}. In a conventional $d$-dimensional TI, topological states appear at $(d-1)$-dimensional boundaries. By contrast, an HOTI hosts boundary modes at higher codimension. In two dimensions, this corresponds to zero-dimensional (0D) states localized at the corners of the sample. Experimental signatures of such corner states have been reported in artificial 2D systems realizing quadrupole topological phases, including mechanical and microwave metamaterials~\cite{SerraGarcia2018,Peterson2018}, as well as in second-order photonic crystals~\cite{Chen2019PRL}. In semiconductor heterostructures, higher-order topological phases associated with double band inversion have been proposed for double HgTe/CdHgTe and three-layer InAs/Ga(In)Sb QWs~\cite{Krishtopenko2021}, building on the established multiband regimes of these systems~\cite{Krishtopenko2016SciRep,Krishtopenko2025PRB,Krishtopenko2018PRBInAs,Krishtopenko2019JETPL,Krishtopenko2019PRB,Avogadri2022PRR}.

The possibility of generating corner states in quantum spin Hall systems by an external magnetic field has attracted considerable attention. In several works it was argued that applying an in-plane magnetic field to a 2D TI may drive the system into a second-order topological phase with states localized at the sample corners~\cite{Ezawa2018,Ren2020,Chen2020PRL}. From the edge-state perspective, the magnetic field gaps the 1D helical spectrum through the Zeeman coupling and produces an effective edge mass. This mass depends on both the crystallographic orientation of the edge and the direction of the magnetic field. Therefore, when the edge direction changes abruptly, for example at a lithographically defined kink, the corresponding mass term changes as well, and a localized in-gap state may appear at the kink. However, the mere presence of a discrete level inside the magnetic-field-induced edge gap is not sufficient to identify it as a HOTI mode protected by a stable bulk invariant. It was recently shown that, in realistic 2D TIs based on HgTe/CdHgTe and InAs/GaInSb/InAs QWs, such magnetic-field-induced corner states should in general be understood as bound states of the effective edge theory, controlled by the relative configuration of magnetic-field-induced edge mass vectors, rather than as topologically protected HOTI corner modes~\cite{KrishtopenkoTeppe2026PRB}. At the same time, the absence of topological protection in this strict sense does not preclude their spectral robustness against weak local perturbations, including short-range static disorder~\cite{Krishtopenko2020PRBDisorder,Krishtopenko2022PRBDisorder,Krishtopenko2025PRR}, as long as the corresponding levels remain isolated inside the edge gap. Together with the possibility of controlling their energy, localization length, and existence condition by the edge geometry and magnetic-field orientation, this makes magnetic-field-induced corner states a promising basis for tunable zero-dimensional levels in semiconductor heterostructures.

\begin{figure}
\includegraphics[width=0.9\columnwidth]{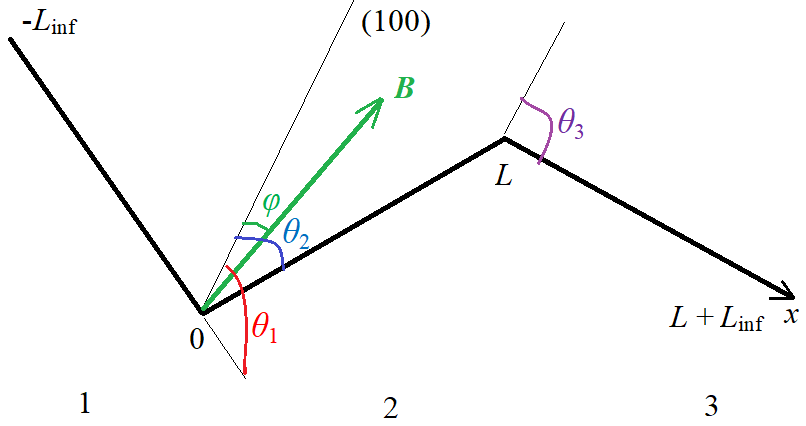}
\caption{\label{Fig:1} Schematic of a HgTe/CdHgTe QW edge containing two lithographically defined kinks, which divide the edge into regions 1, 2, and 3. The orientations of the three edge segments are specified by the angles $\theta_{1,2,3}$ measured with respect to the crystallographic direction (100). The central segment lies along the $x$ direction and extends from $x=0$ to $x=L$. The in-plane magnetic field $\mathbf{B}$ forms an angle $\phi$ with the (100) direction. The full edge is truncated at a length $L_{\rm inf}\gg L$.}
\end{figure}

In the present work we go beyond the existence problem for a single magnetic-field-induced corner state and consider a 2D TI whose edge contains two lithographically defined kinks, as schematically shown in Fig.~\ref{Fig:1}. The two corresponding in-gap states, localized inside the magnetic-field-induced gap of the 1D edge spectrum, can form an effective two-level subsystem. Unlike conventional electrostatically confined levels, this subsystem is defined by the edge geometry and by the magnetic-field-induced edge masses, and its parameters can be tuned by the kink separation and by the orientation of the in-plane magnetic field. This provides a simple setting for studying coherent electric-field-driven dynamics between two localized edge-state levels.

Electrical control of a two-level spin system is one of the central ingredients of semiconductor spin-qubit architectures~\cite{Burkard2023}. In semiconductor nanostructures, coherent spin manipulation can be achieved through electric-dipole spin resonance (EDSR), where spin-orbit coupling converts an oscillating electric perturbation into an effective spin-driving mechanism~\cite{RashbaEfros2003,GolovachLoss2006,Tarucha2006}. Electrically driven coherent spin rotations have also been demonstrated experimentally in gated semiconductor quantum dots and strong-spin-orbit-coupled semiconductor nanostructures~\cite{Nowack2007,NadjPerge2010}. Since spin-momentum coupling is an intrinsic property of helical edge states in 2D topological insulators, localized edge states provide a natural setting in which electrically driven population transfer can be accompanied by a controlled evolution of the spin degree of freedom. This motivates us to investigate whether a pair of magnetic-field-induced corner states can operate as an electrically addressable two-level subsystem.

Related ideas of electrically controlled dynamics of localized states at the edge of a TI have been considered previously for models with  a quantum dot defined by magnetic barriers ~\cite{jetp20}. From the perspective of possible qubit-oriented architectures, the key requirement is the availability of two isolated and controllable discrete levels between which resonant transitions can be driven with weak leakage into continuum states. A conceptually related proposal was also put forward for a HgTe/CdHgTe QW patterned by a periodic array of holes, where a pair of discrete levels appears inside the minigap of the superlattice spectrum~\cite{Kachorovskii2023}. In contrast to such a superlattice-based realization, the system considered below forms the two localized levels directly at lithographically defined kinks of a single edge. Their properties are therefore controlled by the edge geometry and by the orientation of the in-plane magnetic field, rather than by an extended periodic modulation.

The main goal of this work is to go beyond the spectral problem of forming magnetic-field-induced corner states and to analyze their driven dynamics in the presence of fully propagating 1D edge states. We consider a double-kink edge geometry in which two localized in-gap states form inside the magnetic-field-induced edge gap. This geometry does not merely provide a pair of discrete levels; it also creates a controlled setting in which coherent resonant transfer, electric-dipole coupling, and leakage into the continuum can be treated within the same effective edge-state framework. We determine the parameter range, in which two localized states exist, calculate their localization properties and dipole matrix elements, and then solve the time-dependent problem under resonant electric driving. 

The resulting dynamics is governed by a competition between coherent population transfer and leakage, induced by the propagating 1D edge states. A stronger electric field increases the Rabi frequency and shortens the manipulation time, but also enhances population loss from the effective two-level subsystem~\cite{jetp20}. A weaker field suppresses leakage at the cost of a longer Rabi period. The double-kink geometry therefore provides a concrete platform for studying the trade-off between fast electric control and the preservation of a two-level regime. The paper is organized as follows. In Sec.~\ref{Sec:Model} we formulate the model and analyze the formation of two discrete localized levels. In Sec.~\ref{Sec:Continuum} we construct the propagating 1D edge states and evaluate the relevant transition matrix elements. In Sec.~\ref{Sec:Dynamics} we study the resonant dynamics, including Rabi oscillations and leakage into the propagating 1D edge states. Finally, Sec.~\ref{Sec:Conclusions} summarizes the main results.

\section{\label{Sec:Model} Hamiltonian and spectrum of localized states}
Let us consider a 2D TI based on a HgTe/CdHgTe QW whose edge contains two kinks, as schematically shown in Fig.~\ref{Fig:1}. The coordinate $x$ is chosen along the bent edge such that the first kink is located at $x=0$ and the second one at $x=L$. The edge segment between the two kinks defines region 2, whereas the two extended outer segments correspond to regions 1 and 3. The crystallographic orientation of the edge is described by the angle $\theta=\theta(x)$ measured with respect to the (100) direction. In the geometry considered below, $\theta(x)$ is a piecewise constant function taking the values $\theta_1$, $\theta_2$, and $\theta_3$ in regions 1, 2, and 3, respectively. A static magnetic field $\mathbf{B}$ is applied in the plane of the QW and forms an angle $\phi$ with the (100) direction. Therefore, the change of edge orientation at $x=0$ and $x=L$ leads to a change of the magnetic-field-induced mass terms in the effective edge Hamiltonian, which may produce localized states at the two kinks. The outer edge segments are assumed to be sufficiently long, so that the total computational length $L_{\rm inf}$ satisfies $L_{\rm inf}\gg L$ and is used below for the normalization of wave functions.

\subsection{Low-energy model for the edge states}
To describe the low-energy edge states of a 2D TI in a static magnetic field, we use the effective edge Hamiltonian~\cite{KrishtopenkoTeppe2026PRB}
\begin{equation}
H_0=\hbar v_F k_x \sigma_z + M_x(x) \sigma_x +M_y(x) \sigma_y .
\label{hedge}
\end{equation}
Here the first term describes the gapless helical edge states of the BHZ model~\cite{Bernevig2006}, while the second and third terms are magnetic-field-induced mass terms generated by the Zeeman coupling to an in-plane magnetic field. The Pauli matrices $\sigma_i$ act in the space of the two counterpropagating edge modes. For the geometry considered in Fig.~\ref{Fig:1}, the magnetic field lies in the QW plane and forms an angle $\phi$ with the crystallographic direction (100). The mass terms are then written as
\begin{equation*} 
M_x(x)=\frac{E_Z}{2}\left[
g_{+} \cos (\theta(x)-\phi) +  g_{-} \cos (3 \theta(x)+\phi)
\right],
\end{equation*}
\begin{equation} 
M_y(x)=\frac{E_Z}{2}\left[
g_{+} \sin (\theta(x)-\phi) +  g_{-} \sin (3 \theta(x)+\phi)
\right],
\label{mxmy}
\end{equation}
where $E_Z=g_1 \mu_B B/2$ is the Zeeman energy scale and $g_{\pm}=1 \pm g_2/g_1$. The parameters $g_1$ and $g_2$ are combinations of the components of the edge-state $g$-factor tensor. Since $\theta(x)$ is piecewise constant, the mass vector $(M_x,M_y)$ is also piecewise constant and changes abruptly at the two kinks.

The third angular harmonic in the terms proportional to $g_-$ originates from the $4\theta$ crystallographic anisotropy of the in-plane edge-state $g$-factor tensor~\cite{KrishtopenkoTeppe2026PRB}: combined with the field orientation $\theta-\phi$ in the local edge frame, it gives $4\theta-(\theta-\phi)=3\theta+\phi$. Neglecting the contributions related to the lack of inversion symmetry in the bulk unit cell of HgTe and CdHgTe, namely bulk inversion asymmetry (BIA)~\cite{Konig2008JPSJ}, as well as possible interface inversion asymmetry (IIA) of HgTe/CdHgTe heterointerfaces~\cite{DurnevTarasenko2016PRB}, the parameters of the edge Hamiltonian take a simple form. This approximation is justified for realistic HgTe/CdHgTe QWs, where the BIA and IIA terms are small~\cite{Krishtopenko2020PRB,Ruffenach2026PRB}. In this case one obtains~\cite{KrishtopenkoTeppe2026PRB}
\begin{equation}
\hbar v_F={\mathbb A}\sqrt{1-\eta^2},
\end{equation}
and
\begin{eqnarray}
g_1=g_e^{(\parallel)}\frac{1-\eta}{2}+ g_h^{(\parallel)}\frac{1+\eta}{2},
\nonumber\\
g_2=g_e^{(\parallel)}\frac{1-\eta}{2}- g_h^{(\parallel)}\frac{1+\eta}{2},
\label{struct}
\end{eqnarray}
where $\eta={\mathbb D}/{\mathbb B}$. Here $g_e^{(\parallel)}$ and $g_h^{(\parallel)}$ are the in-plane $g$ factors of the $E1$ and $H1$ subbands, respectively. They originate from the contributions of the bulk $\Gamma_6$, $\Gamma_8$, and $\Gamma_7$ bands~\cite{Winkler2003}. The parameters ${\mathbb A}$, ${\mathbb B}$, ${\mathbb D}$, $g_e^{(\parallel)}$, and $g_h^{(\parallel)}$ depend on the QW width and on the Cd content in the CdHgTe barriers. In the numerical calculations below we use ${\mathbb A}=365$~meV$\cdot$nm, ${\mathbb B}=-768$~meV$\cdot$nm$^2$, ${\mathbb D}=-593$~meV$\cdot$nm$^2$, $g_e^{(\parallel)}=27$, and $g_h^{(\parallel)}=-1.2$. These values are characteristic of a $7$~nm-wide HgTe/Cd$_{0.7}$Hg$_{0.3}$Te QW grown on a (001)-oriented CdTe buffer~\cite{Krishtopenko2018PRB}. For the same prototype QW, the BHZ mass parameter is $\mathbb M=-4.53$ meV, corresponding to a 2D bulk gap $E_{g}=2|M|\simeq9.1$ meV~\cite{Krishtopenko2018PRB}. For this prototype QW, Eqs.~(\ref{struct}) yield
\begin{equation}
g_1=2.012, \quad g_2=4.139.
\label{gfac}
\end{equation}

\begin{figure}
\includegraphics[width=0.8\columnwidth]{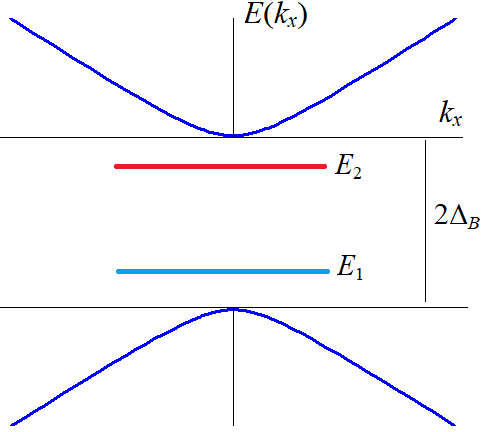}
\caption{\label{FigSpectrum} Schematic of the spectrum consisting of the upper and lower continuum branches described by Eq.~(\ref{encont}). The blue curves represent the corresponding continuum dispersions bordering the common magnetic-field-induced gap $2\Delta_B$. For the double-kink geometry shown in Fig.~\ref{Fig:1}, two discrete levels $E_{1,2}$ may appear inside this gap for suitable parameters. The corresponding wave functions are localized at the kinks of the edge.}
\end{figure}

The eigenfunctions of Hamiltonian~(\ref{hedge}) are sought in the form of two-component spinors
\begin{equation}
\psi_x=
\left(
\begin{array}{c}
a \\
b
\end{array}
\right)
e^{\lambda x},
\label{wf}
\end{equation}
where the parameter $\lambda$ can be either purely imaginary or real. In the former case, the solutions describe propagating edge states with plane-wave factors $\exp(\pm i k_x x)$. For a given region $n=1,2,3$, where the edge orientation is fixed and the mass terms are constant, Hamiltonian~(\ref{hedge}) gives the continuum spectrum schematically shown in Fig.~\ref{FigSpectrum},
\begin{equation}
\varepsilon_{1,2}(k_x(n))=\pm \sqrt{k_x(n)^2+|M_n|^2}.
\label{encont}
\end{equation}
Here $\varepsilon_{1,2}(k_x(n))=E_{1,2}(k_x(n))/(\hbar v_F)$ is the energy rescaled by $\hbar v_F$, and the index $n$ labels the three spatial regions shown in Fig.~\ref{Fig:1}. Thus, a state with the same rescaled energy $\varepsilon$ generally corresponds to different longitudinal wave vectors $k_x(n)$ in different regions.

The in-plane magnetic field opens a common gap $2\Delta_B$ in the edge-state spectrum with
\begin{equation}
\Delta_B=\min_{n=1,2,3} |\hbar v_F M_n|,
\label{deltab}
\end{equation}
where
\begin{equation}
|M_n|^2=\frac{(M_x(n))^2+(M_y(n))^2}{(\hbar v_F)^2}.
\label{mn}
\end{equation}
The quantities $M_{x,y}(n)$ are obtained from Eq.~(\ref{mxmy}) by setting $\theta=\theta_n$ in the corresponding region. The discrete localized levels considered below lie inside this common magnetic-field-induced gap. For energies outside this interval, at least one of the three edge regions supports propagating continuum states. The corresponding propagating solutions described by the spectrum~(\ref{encont}) will be considered in the next section. Here we focus on localized states, for which the parameter $\lambda$ in Eq.~(\ref{wf}) is real and the wave functions decay away from the edge kinks, as illustrated in Fig.~\ref{FigSpectrum}.

The boundary conditions for localized states are given by the continuity of the wave function at the two kinks, $x=0$ and $x=L$, together with the decay of the wave function at infinity:
\begin{equation}
\left\{
\begin{array}{c}
\Psi(x\to \pm \infty)=0, 
\\ 
\Psi(x+0)=\Psi(x-0), \quad \Psi(L+0)=\Psi(L-0).
\end{array}
\right.
\label{bc}
\end{equation}
These conditions are satisfied by the following piecewise-defined solutions. In region 1, for $x<0$, the wave function is written as
\begin{equation}
\Psi_1(x)=A
\left(
\begin{array}{c}
1 \\ 
\dfrac{\sqrt{|M_1|^2-\varepsilon^2}-i \varepsilon}{M_{+1}}
\end{array}
\right)
e^{\sqrt{|M_1|^2-\varepsilon^2}\,x},
\label{wfloc1}
\end{equation}
where
\begin{equation}
M_{\pm n}=\frac{M_x(n) \pm i M_y(n)}{\hbar v_F}.
\label{mpm}
\end{equation}
The coefficient $A$ is determined from the boundary conditions~(\ref{bc}). 

In region 2, for $0<x<L$, the localized-state solution contains both increasing and decreasing exponential components,
\begin{eqnarray}
\Psi_2(x)&=&B_1
\left(
\begin{array}{c}
1 \\ 
\dfrac{\sqrt{|M_2|^2-\varepsilon^2}-i \varepsilon}{M_{+2}}
\end{array}
\right)
e^{\sqrt{|M_2|^2-\varepsilon^2}\,x}
\nonumber\\
&&+B_2\left(
\begin{array}{c}
1 \\ 
\dfrac{-\sqrt{|M_2|^2-\varepsilon^2}-i \varepsilon}{M_{+2}}
\end{array}
\right)
e^{-\sqrt{|M_2|^2-\varepsilon^2}\,x}.
\label{wfloc2}
\end{eqnarray}
In region 3, for $x>L$, the decaying solution has the form
\begin{equation}
\Psi_3(x)=C
\left(
\begin{array}{c}
1 \\ 
\dfrac{-\sqrt{|M_3|^2-\varepsilon^2}-i \varepsilon}{M_{+3}}
\end{array}
\right)
e^{-\sqrt{|M_3|^2-\varepsilon^2}\,x}.
\label{wfloc3}
\end{equation}

The coefficients $A$, $B_1$, $B_2$, and $C$ are found from the boundary conditions~(\ref{bc}). Substituting Eqs.~(\ref{wfloc1})--(\ref{wfloc3}) into these conditions gives a homogeneous system of linear equations for the four coefficients. After solving this system, the resulting wave functions are normalized on the full $x$ axis. Nontrivial solutions exist only at rescaled energies $\varepsilon_i$ satisfying the characteristic equation
\begin{equation}
\Delta(\varepsilon_i)=0,
\label{chareq}
\end{equation}
where $\Delta$ is the determinant of the coefficient matrix. Equation~(\ref{chareq}) is solved numerically for different sets of parameters $\theta_{1,2,3}$, $\phi$, and $L$. In the following subsection, we identify the regions of this parameter space where Eq.~(\ref{chareq}) has two rescaled-energy solutions $\varepsilon_{1,2}$ corresponding to two localized states that can be resonantly coupled by a time-periodic electric field.

\subsection{Numerical results for localized states}
The analysis of the characteristic equation~\eqref{chareq} shows that the broadest parameter region in which two discrete levels exist inside the gap shown in Fig.~\ref{FigSpectrum} is obtained for the symmetric choice $\theta_3=\theta_1$ in Fig.~\ref{Fig:1}. For definiteness, we set $\theta_2=0$ in the following and, at fixed magnetic-field amplitude, perform the search for solutions of Eq.~(\ref{chareq}) in the $(\theta_1,\phi)$ parameter plane.

In the numerical calculations we choose $E_Z=150~\mu$eV. With the edge-state $g$ factors given in Eq.~\eqref{gfac}, this value corresponds to an in-plane magnetic field $B=1.6$~T, which is readily accessible in experiments on 2D TIs. We consider a structure with the length of the central edge segment $L=250$~nm. This corresponds to a mesoscopic sample size compatible with realistic lithographic fabrication constraints.

\begin{figure}
\includegraphics[width=0.9\columnwidth]{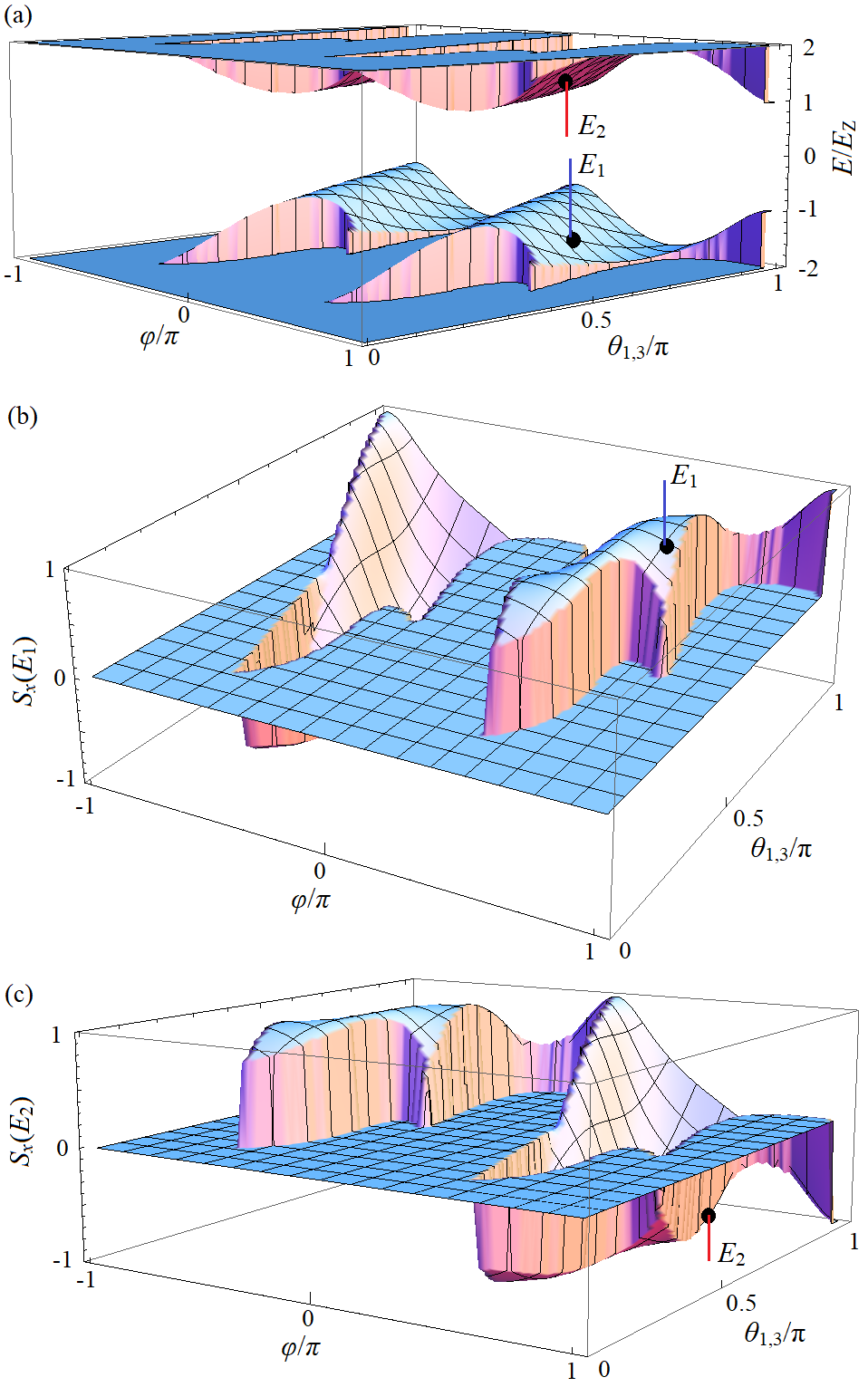}
\caption{\label{FigEnThetaPhi} (a) Spectrum of the two discrete levels $E_{1,2}$ shown schematically in Fig.~\ref{FigSpectrum} as a function of the angles $\theta_1=\theta_3$ and $\phi$ defined in Fig.~\ref{Fig:1}. The other parameters are $\theta_2=0$, $L=250$~nm, and $E_Z=150~\mu$eV. The uniform background regions at the top and bottom correspond to the absence of a pair of localized levels. The selected levels $E_{1,2}=\mp 1.342\,E_Z$ at $\theta_1=\theta_3=2\pi/3$ and $\phi=\pi/2$ are marked by black circles. (b), (c) Distributions of the $x$ component of spin for the lower (b) and upper (c) levels shown in panel~(a). For the selected pair of levels, the $x$-spin projections have opposite signs and nearly equal magnitudes.
}
\end{figure}

An example of the calculated discrete spectrum in the $(\theta_1,\phi)$ parameter plane is shown in Fig.~\ref{FigEnThetaPhi}. Panel~(a) presents the energies of the two discrete levels in units of $E_Z$. The uniform background regions at the top and bottom of the panel correspond to parameter values for which a pair of localized levels is not formed. The black circles mark the levels $E_1=-1.342\,E_Z$ and $E_2=1.342\,E_Z$, obtained for $\theta_1=\theta_3=2\pi/3$ and $\phi=\pi/2$. For the chosen parameters, $\Delta_B=1.345\,E_Z$. Thus, the selected levels $E_{1,2}=\mp 1.342E_Z$ lie very close to the boundaries of the commonmagnetic-field-induced edge gap, with $\Delta_B-|E_{1,2}|=0.003\,E_Z$. This proximity to the continuum implies that leakage into the extended edge states may become relevant during the driven dynamics considered below. 
This pair of levels is used in most of the calculations below. The choice of these parameters is also motivated by the spin properties of the corresponding states. Panels~(b) and (c) of Fig.~\ref{FigEnThetaPhi} show the $x$ component of spin, $S_x$, in units of $\hbar/2$, for the lower and upper levels in panel~(a). For the selected pair of states, the values of $S_x$ are nearly equal in magnitude and opposite in sign. This property is useful for associating the two localized levels with opposite projections of spin along the $x$ direction in possible applications of the structure as a controllable two-level subsystem.

For the pair of levels chosen in Fig.~\ref{FigEnThetaPhi}, the absolute value of the dipole matrix element
\begin{equation}
X_{12}=\langle \Psi(E_1)|x|\Psi(E_2)\rangle.
\label{x12}
\end{equation}
Here, $\Psi(E_1)$ and $\Psi(E_2)$ denote the complete normalized eigenfunctions corresponding to the two discrete levels $E_1$ and $E_2$ marked in Fig.~\ref{FigEnThetaPhi}(a). They should not be confused with the region-resolved wave functions $\Psi_n(x)$, with $n=1,2,3$, introduced in Eqs.~\eqref{wfloc1}--\eqref{wfloc3}.
For the selected pair of levels, the absolute value of the dipole matrix element is $|X_{12}|=885~{\rm nm}\simeq 3.5L$. Its magnitude is sufficiently large for efficient electric driving: for the field amplitudes considered below, the corresponding coupling produces Rabi oscillations with frequencies in the range $20$--$40~{\rm GHz}$, while the leakage into the continuum can be kept small in the weak-driving regime.

The coordinate $x$ entering Eq.~\eqref{x12} is the curvilinear coordinate measured along the 1D edge channel. This choice is motivated by the transport geometry relevant to experimental studies of QSH edge states, where the directly accessible quantity is the voltage drop between contacts connected by a selected edge channel~\cite{Nowack2013,Spanton2014,Benlemqwanssa2024}. In multi-probe transport measurements, the edge contribution can be separated from the residual bulk contribution, allowing the voltage dependence of transport along a particular edge channel to be analyzed independently. In this representation, the natural longitudinal driving scale for a curved edge is therefore the voltage drop divided by the length measured along the channel, rather than the projection of a spatially uniform 2D electric field onto a fixed crystallographic direction.

\begin{figure}
\includegraphics[width=1.0\columnwidth]{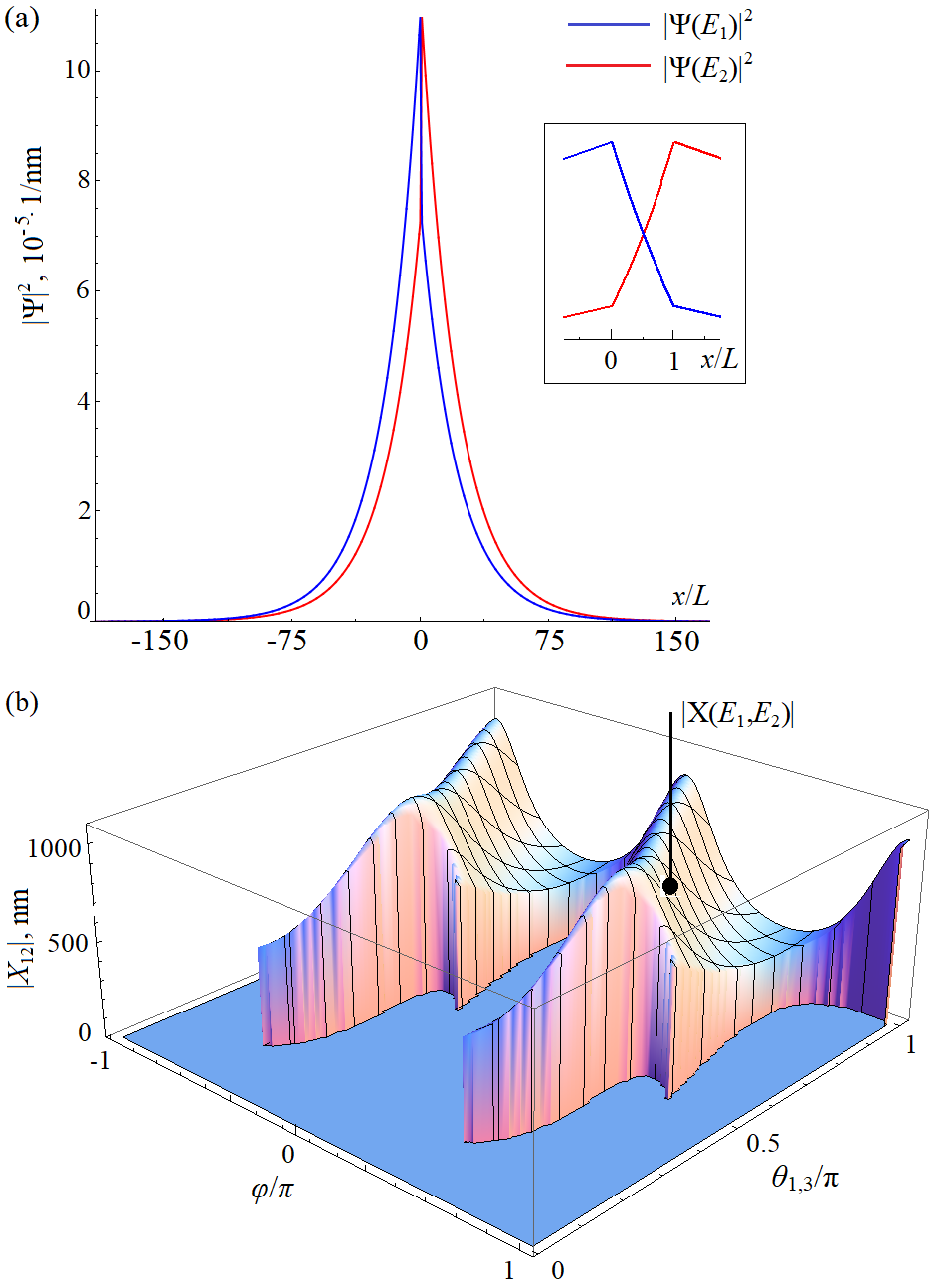}
\caption{\label{FigPsiX12}
(a) Probability-density distributions $|\Psi|^2$ for the levels$E_{1,2}$ marked in Fig.~\ref{FigEnThetaPhi}, shown on a linear scale. The extended spatial range demonstrates the exponential localization of both states. For the selected parameters, the probability-density tails have a characteristic decay length of order $\ell_{\rm loc}\sim 30L$. Inset: the probability maxima are located predominantly near the left and right ends of the central segment $0<x<L$. (b) Absolute value of the dipole matrix element~(\ref{x12}) as a function of the angles $\theta_1=\theta_3$ and $\phi$. The marked value $|X_{12}|=885$~nm corresponds to the selected pair of discrete levels whose probability-density distributions are shown in panel~(a).}
\end{figure}

Figure~\ref{FigPsiX12}(a) shows the probability-density distributions for the states $\Psi(E_1)$ and $\Psi(E_2)$ corresponding to the lower and upper discrete levels in Figs.~\ref{FigSpectrum} and~\ref{FigEnThetaPhi}. Figure~\ref{FigPsiX12}(b) presents the absolute value of the dipole matrix element~\eqref{x12} over the full parameter plane $(\theta_1=\theta_3,\phi)$. As seen from the inset in Fig.~\ref{FigPsiX12}(a), the probability maxima for the lower and upper discrete states are localized predominantly near the left kink at $x=0$ and the right kink at $x=L$, respectively. Since the selected levels lie very close to the continuum thresholds, their localization lengths are much larger than $L$. To quantify their spatial extent, we characterize the localization by the exponential decay length of the probability density. For a wave-function tail proportional to $\exp(-|\lambda|x)$, the probability density decays as $\exp(-2|\lambda|x)$, corresponding to $\ell_{\rm loc}=1/(2|\lambda|)$. For the selected states, the slowest exponential tails give a characteristic scale of order $\ell_{\rm loc}\sim 30L\simeq 7.5~\mu{\rm m}$. Equivalently, the corresponding wave-function amplitude decay length is $1/|\lambda|\sim60L\simeq15~\mu{\rm m}$, which is much larger than $L=250$ nm. The absence of nodes in the probability densities is not anomalous, since the localized states are two-component Dirac spinors and the usual nodal argument for a scalar one-dimensional Schr\"odinger problem does not apply.

In a realistic device, static disorder and edge imperfections can modify both the energies and localization properties of the corner states. Within the effective 1D description, such perturbations may be represented by a random scalar potential and by fluctuations of the magnetic mass terms. Weak disorder therefore shifts $E_{1,2}$ and renormalizes the spatial profiles and localization lengths of the states. Nevertheless, as discussed in Ref.~\cite{KrishtopenkoTeppe2026PRB}, magnetic-field-induced corner states may remain spectrally identifiable in-gap excitations under weak perturbations. The effective two-level description is therefore expected to remain applicable as long as the edge gap stays open and the two localized levels remain isolated from the continuum.

\section{\label{Sec:Continuum} Continuum states and transition matrix elements}
Although the main focus of this work is on the states localized at the kinks of the edge in Fig.~\ref{Fig:1}, the propagating 1D states can also participate in the driven dynamics and provide a leakage channel out of the effective two-level subsystem. Under a time-periodic electric field, these states can be coupled to the localized levels and thus provide a leakage channel out of the effective two-level subsystem. In the double-kink geometry, propagating continuum states are partially reflected and transmitted at the two kinks, $x=0$ and $x=L$. Therefore, their wave functions cannot be represented by a single expression of the form~(\ref{wf}) over the entire edge. Instead, they are constructed piecewise in regions 1, 2, and 3, in analogy with the localized states~(\ref{wfloc1})--(\ref{wfloc3}), but with purely imaginary exponents. For a fixed energy in the spectrum of Eq.~(\ref{encont}), there are two families of propagating states: states $\Psi^{(R)}$ incident from the left and propagating from left to right, and states $\Psi^{(L)}$ incident from the right and propagating from right to left. Both families considered below are constructed for energies at which propagating solutions exist in all three edge regions.

Let us first consider the states $\Psi^{(R)}$ corresponding to a wave incident from the left. In region 1 of Fig.~\ref{Fig:1}, for $x<0$, the wave function has the form
\begin{equation}
\Psi^{(R)}_1(x)=
\left( 
\begin{array}{c}
1 \\ 
\alpha_{+1}
\end{array}
\right)
e^{i k_1 x} +
A_{-1}\left( 
\begin{array}{c}
1 \\ 
\alpha_{-1}
\end{array}
\right)
e^{-i k_1 x}.
\label{wfrm1}
\end{equation}
The first term in Eq.~(\ref{wfrm1}) describes the incident wave with unit amplitude, while the second term describes the wave reflected from the double-kink region. Here we introduced
\begin{equation}
\alpha_{\pm n}=\frac{\varepsilon \mp \sqrt{\varepsilon^2-|M_n|^2}}{M_{-n}},
\quad
k_n=\sqrt{\varepsilon^2 - |M_n|^2},
\label{alphankn}
\end{equation}
where $M_n$ and $M_{\pm n}$ are defined in Eqs.~(\ref{mn}) and~(\ref{mpm}), respectively.

In region 2, for $0<x<L$, the wave function contains two components propagating in opposite directions,
\begin{equation}
\Psi^{(R)}_2(x)=
A_{+2}\left( 
\begin{array}{c}
1 \\ 
\alpha_{+2}
\end{array}
\right)
e^{i k_2 x} +
A_{-2}\left( 
\begin{array}{c}
1 \\ 
\alpha_{-2}
\end{array}
\right)
e^{-i k_2 x}.   
\label{wfrm2}
\end{equation}
In region 3, for $x>L$, only the transmitted right-moving wave remains,
\begin{equation}
\Psi^{(R)}_3(x)=
A_{+3}\left( 
\begin{array}{c}
1 \\ 
\alpha_{+3}
\end{array}
\right)
e^{i k_3 x}.  
\label{wfrm3}
\end{equation}
Propagating 1D solutions exist simultaneously in all three edge regions for $|E|>\Delta_B^{(1)}$, where
\begin{equation}
\Delta_B^{(1)}=\max_{n=1,2,3}|\hbar v_F M_n|.
\label{deltab1}
\end{equation}
For the parameters used in the dynamical calculations, $\Delta_B^{(1)}=2.057E_Z\simeq0.309~{\rm meV}$.

The coefficients $A_{-1}$, $A_{\pm2}$, and $A_{+3}$ are determined from the continuity of the wave function at $x=0$ and $x=L$, i.e., from the second and third conditions in Eq.~(\ref{bc}). Substitution of Eqs.~(\ref{wfrm1})--(\ref{wfrm3}) gives a system of four inhomogeneous linear equations, where the inhomogeneous term originates from the incident wave of unit amplitude in Eq.~(\ref{wfrm1}). This system has a solution for any incident-wave energy satisfying $|E|>\Delta_B^{(1)}$. After the coefficients are found, the continuum wave functions are normalized to unity on a large but finite interval of length of order $L_{\rm inf}$, with $L_{\rm inf}\gg L$.

The second family of continuum states, $\Psi^{(L)}$, corresponds to waves incident from the right and propagating along the edge from right to left. These solutions are constructed in the same way as the states $\Psi^{(R)}$. The difference is that the incident wave with unit amplitude is proportional to $e^{-i k_3 x}$ and enters the solution in region 3, whereas in region 1 only the transmitted left-moving wave proportional to $e^{-i k_1 x}$ remains. The corresponding coefficients are again obtained from the continuity conditions at $x=0$ and $x=L$, and the normalization is performed in the same way. In the dynamical calculations below, continuum states of both families, $\Psi^{(R)}$ and $\Psi^{(L)}$, are included.

For the numerical simulations of the dynamics discussed in the next section, the propagating 1D states are discretized in energy. In the construction of the propagating continuum wave functions, it is convenient to use the rescaled energy $\varepsilon=E/(\hbar v_F)$, which has the dimension of inverse length. However, the energy window used in the dynamical calculations is specified in terms of the physical energy $E$. Our calculations show that it is sufficient to include continuum states with physical energies in the interval $|E|<n_b\Delta_B$, where $n_b\sim 200$. Following our previous treatment of continuum-assisted dynamics for a quantum dot at the edge of a TI~\cite{jetp20}, we introduce a uniform discretization of this energy interval with the step
\begin{equation}
\delta E = \frac{n_b \Delta_B}{N},
\end{equation}
where $N\sim 1000$. This discretization provides an adequate description of population leakage from the two-level subsystem into the continuum, caused by nonzero dipole matrix elements
\begin{equation}
X_{nE}^{(R,L)}=\langle \Psi_n|x|\Psi_E^{(R,L)}\rangle.
\label{xnc}
\end{equation}
Here, $\Psi_n$ with $n=1,2$ denotes one of the two localized states, while $\Psi_E^{(R,L)}$ is a propagating 1D state belonging to either the right-incident or left-incident family. The same curvilinear coordinate $x$ is used in Eq.~\eqref{xnc} as in Eq.~\eqref{x12}. The dependence on the real-space edge configuration, including $\theta_{1,2,3}$ and $\phi$, enters through the localized and continuum wave functions determined by the corresponding piecewise mass vectors.

\begin{figure}
\includegraphics[width=1.0\columnwidth]{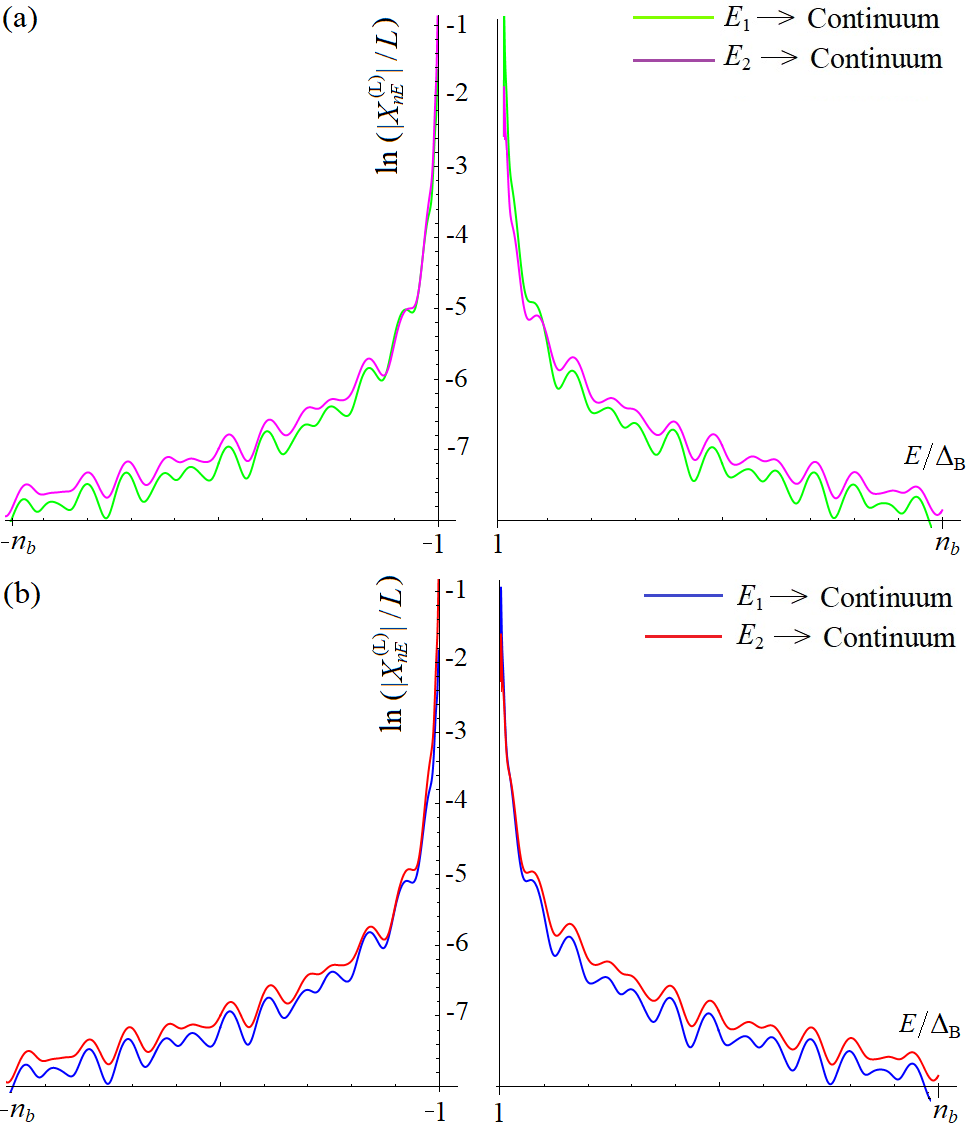}
\caption{\label{FigXnmCont} Logarithm $\ln(|X_{nE}^{(R,L)}|/L)$ of the absolute value of the normalized dipole matrix element in Eq.~\eqref{xnc} for transitions into continuum states of the families (a)~$\Psi^{(L)}$ and (b)~$\Psi^{(R)}$. Each panel shows transitions from both localized levels $E_1$ and $E_2$. The lower and upper propagating 1D branches correspond to $-n_b\Delta_B<E<-\Delta_B$ and $\Delta_B<E<n_b\Delta_B$, respectively, with $n_b=200$. For the selected pair of localized levels, $\Delta_B=1.345\,E_Z\simeq 0.202~{\rm meV}$.}
\end{figure}

In practice, the 1D states that contribute appreciably to the dynamics occupy a much narrower energy range than the full interval $|E|<n_b\Delta_B$. This is illustrated in Fig.~\ref{FigXnmCont}, which shows $\ln(|X_{nE}^{(R,L)}|/L)$ as a function of the normalized continuum energy $E/\Delta_B$. Panel~(a) corresponds to the family $\Psi^{(L)}$, whereas panel~(b) corresponds to the family $\Psi^{(R)}$. In each case, the matrix elements for transitions from both discrete levels $E_1$ and $E_2$ are shown. The lower and upper propagating 1D branches correspond to $-n_b\Delta_B<E<-\Delta_B$ and $\Delta_B<E<n_b\Delta_B$, respectively, with $n_b=200$.

As seen from Fig.~\ref{FigXnmCont}, the matrix elements for transitions into the lower and upper continuum branches exhibit similar energy dependences and rapidly decrease away from the edges of the magnetic-field-induced gap. Only a part of the chosen energy window gives an appreciable contribution. The largest matrix elements occur near the continuum threshold; for the parameters considered here, the relevant transitions involve continuum states within several meV from the gap edges. Nevertheless, in the dynamical calculations below we retain all propagating 1D states with physical energies in the intervals $[-n_b\Delta_B,-\Delta_B]$ and $[\Delta_B,n_b\Delta_B]$. For $n_b=200$, this corresponds to an energy window of the order of $[-30,30]$~meV and to approximately $2000$ continuum states included in the calculation.

Extending this energy window further is not useful within the present low-energy edge-state model. At higher energies, the edge-state spectrum starts to overlap with bulk 2D states of the HgTe/CdHgTe QW, against which the edge contribution is no longer well separated~\cite{Krishtopenko2018PRB}. The bulk 2D states in the considered QW lie outside a sufficiently large energy gap that contains the edge-state spectrum~(\ref{encont}). Moreover, the rapid decrease of the matrix elements~(\ref{xnc}) in Fig.~\ref{FigXnmCont} suggests that transitions to even more distant bulk states should have small amplitudes. Such states are therefore not included in the approximation used below.

We note that the continuum-propagating-state model described above is not unique. In our previous studies of population leakage into the continuum from discrete donor levels~\cite{prb2019} and from a multilevel quantum dot~\cite{prappl2020} in an InSb-based nanowire, the continuum was represented by a large number of closely spaced discrete levels. Each such level corresponded to a state localized in a large interval $[-L_{\rm inf},L_{\rm inf}]$ with distant infinitely high walls. An analogous continuum model can also be constructed for the present system described by Hamiltonian~(\ref{hedge}). Our calculations show that this alternative approach gives results for the time evolution and characteristic leakage times that are close to those obtained with the two-family propagating-state model $\Psi^{(R,L)}$ introduced in this section. This agreement indicates that, for the fully propagating continuum sector considered here, the calculated dynamics is not sensitive to the particular representation of that sector.

\section{\label{Sec:Dynamics} Resonant dynamics in an electric field}
We now turn to the central dynamical problem of this work: resonant electric manipulation of the two magnetic-field-induced localized states in the presence of continuum edge states. The electric-field frequency is chosen to be resonant with the energy separation between the discrete levels $E_{1,2}$ shown in Fig.~\ref{FigSpectrum}. The purpose of this analysis is to examine the driven two-level dynamics when coupling to the fully propagating continuum states constructed in Sec.~\ref{Sec:Continuum} is included explicitly. The time-dependent electric drive is modeled in the quasistatic approximation as a longitudinal AC voltage drop along the 1D edge channel, described by the scalar potential~\cite{jetp20}
\begin{equation}
V(x,t)=-F x \sin \omega t .
\label{velec}
\end{equation}
Here $x$ is the curvilinear coordinate introduced above, and $F$ represents the energy gradient associated with the longitudinal voltage drop along the edge channel. For an edge channel of length $L_{\rm ch}$ subject to an AC voltage with amplitude $V_{\rm ac}$, assuming a uniform longitudinal voltage gradient along the channel gives
\begin{equation}
F=\frac{eV_{\rm ac}}{L_{\rm ch}},
\label{FVac}
\end{equation}
where $e$ is the elementary charge. Thus, the local longitudinal electric field follows the tangent to the 1D edge channel and changes its direction at the kinks together with the channel itself. Equation~\eqref{velec} therefore describes a voltage-biased curved edge channel and should not be interpreted as a spatially uniform electric field applied to the entire 2D QW. Below, $F$ is measured in $\mu{\rm eV}/{\rm nm}$. We choose the resonant frequency corresponding to the transition energy between the two discrete levels, $\hbar\omega=E_2-E_1=0.398$~meV, which gives the linear frequency $f=96.4$~GHz. This frequency range is accessible with modern microwave and millimeter-wave techniques.

\begin{figure}
\includegraphics[width=0.9\columnwidth]{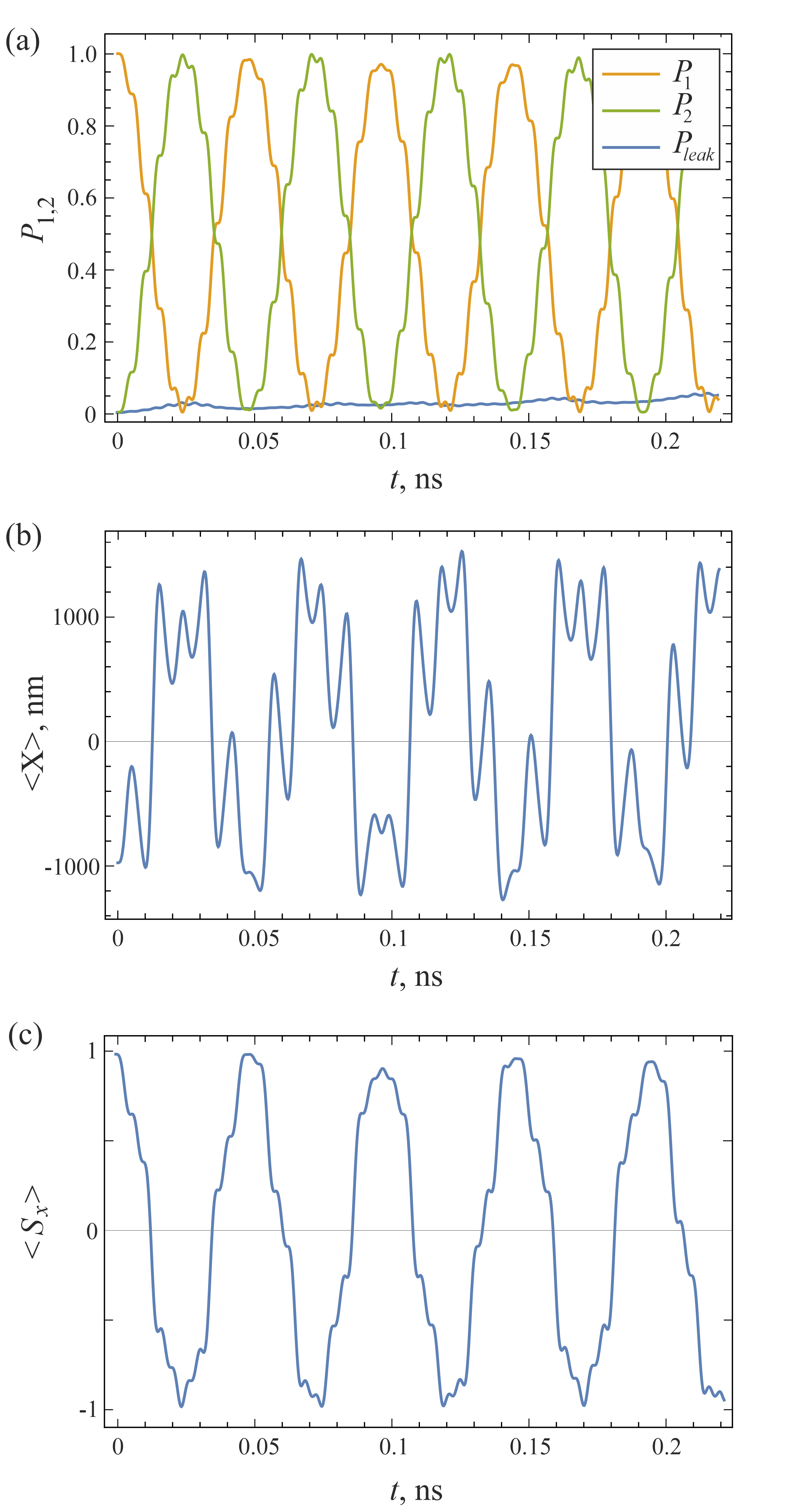}
\caption{\label{FigDynF0p1} Example of the dynamics for the field amplitude $F=0.1~\mu$eV/nm. (a) Populations of the discrete levels, $P_{1,2}(t)=|C_{1,2}(t)|^2$, and leakage into the propagating 1D edge states, $P_{\rm leak}=1-|C_1|^2-|C_2|^2$. The dynamics shows Rabi oscillations with the linear frequency $f_R=40$~GHz and leakage $P_{\rm leak}\sim 0.05$ over a time interval of about $0.02$~ns. (b,c) Time evolution of the expectation values $\langle X(t)\rangle$ and $\langle S_x(t)\rangle$. Both quantities follow the population oscillations in panel~(a) and can be used to characterize the state of the two-level subsystem.}
\end{figure}

The solution of the time-dependent Schr\"{o}dinger equation with the Hamiltonian $H_0+V(x,t)$, where $H_0$ is defined in Eq.~(\ref{hedge}), is expanded over the stationary eigenstates $\Psi_n(x)$ as
\begin{equation}
\Psi(x,t)=\sum_n C_n(t) e^{-i E_n t/\hbar} \Psi_n(x).
\label{psit}
\end{equation}
This basis includes both the two localized states of the discrete spectrum discussed in Sec.~\ref{Sec:Model} and the propagating 1D states constructed in Sec.~\ref{Sec:Continuum}. The expansion coefficients $C_n(t)$ obey the system of ordinary differential equations
\begin{equation}
\frac{dC_n}{dt}=i\frac{F}{\hbar}\sin\omega t\sum_l C_l(t) X_{nl}e^{-i(E_l-E_n)t/\hbar}.
\label{cnt}
\end{equation}
The dynamics is governed by the dipole matrix elements of the coordinate operator, $X_{nl}$, introduced above in Eqs.~(\ref{x12}) and~(\ref{xnc}). As the initial condition, we take the system to be fully populated on the lower discrete level $E_1$. 

In an ideal closed two-level system, the total population remains within the two discrete states, $P_1(t)+P_2(t)=1$, and resonant driving produces essentially single-frequency Rabi oscillations. In the present system, however, the coordinate operator couples the localized states also to the propagating continuum. This produces two qualitative effects absent from the ideal two-level description: population leakage from the discrete subspace and additional coherent frequency components due to the off-resonant admixture of propagating continuum states. The latter give rise to the non-sinusoidal fine structure of the Rabi oscillations discussed below. Increasing the field amplitude enhances the Rabi frequency but also strengthens the coupling to the propagating continuum, whereas reducing it suppresses leakage at the cost of slower oscillations. The numerical analysis below therefore identifies a regime in which the Rabi dynamics remains fast while the continuum contribution stays weak over several oscillation periods.

For a fixed realization of weak static disorder, the shifts of the localized-state energies and wave functions will generally modify the resonance frequency, the dipole matrix elements $X_{12}$ and $X_{nE}$, and hence the Rabi frequency and continuum leakage. Static disorder, however, does not by itself constitute a decoherence mechanism: if two spectrally isolated localized states survive, the driving frequency can be retuned to their actual level separation and coherent Rabi oscillations remain possible with renormalized parameters. Time-dependent fluctuations and the resulting dephasing require a separate treatment beyond the scope of the present work. We start with the driving amplitude $F=0.1~\mu{\rm eV}/{\rm nm}$. Since the selected discrete levels lie close to the continuum thresholds and possess long exponential tails, their dipole coupling to the propagating continuum states is not negligible at this driving amplitude. Therefore, the expansion~\eqref{psit} is performed in an extended basis including both the two discrete levels and the discretized continuum states.

An example of the calculated dynamics for $F=0.1~\mu$eV/nm is shown in Fig.~\ref{FigDynF0p1}. Panel~(a) presents the populations of the two discrete levels, $P_{1,2}(t)=|C_{1,2}(t)|^2$, together with the leakage probability into the propagating gapped 1D edge states,
\begin{equation}
P_{\rm leak}=1-|C_1|^2-|C_2|^2 .
\label{pleak}
\end{equation}
The populations $P_1$ and $P_2$ exhibit Rabi oscillations with the linear frequency
\begin{equation}
f_R=40~{\rm GHz}.
\label{frabi}
\end{equation}

This value corresponds to characteristic manipulation times on the scale of several tens of picoseconds. The populations in Fig.~\ref{FigDynF0p1}(a) exhibit regular high-frequency modulations superimposed on the slower Rabi oscillations. Their origin is related to the multilevel character of the dynamics described by Eq.~\eqref{cnt}: in addition to the resonant coupling between the localized states $E_1$ and $E_2$, the electric perturbation admixes propagating 1D edge states and introduces additional frequency components. Their interference produces the non-sinusoidal fine structure of $P_1(t)$ and $P_2(t)$. This fine structure is therefore a direct manifestation of the continuum-assisted multilevel dynamics and is absent in an isolated two-level model. The most pronounced modulations appear close to the extrema of the slow Rabi envelope, while the dominant oscillatory component remains characterized by the Rabi frequency $f_R=40~{\rm GHz}$. At this stage, the coherent description should therefore be viewed as a first estimate of fast resonant dynamics. The actual applicability of such a two-level subsystem to qubit-oriented architectures requires a separate analysis of spin relaxation and decoherence mechanisms, which in HgTe/CdHgTe structures may depend sensitively on the transport regime and sample parameters~\cite{Minkov2012PRB,Kadykov2018PRL}. 

To quantify the validity of the effective two-level description, it is useful to introduce the total population retained in the discrete subspace,
\begin{equation*}
P_{\rm TLS}(t)=P_1(t)+P_2(t)=1-P_{\rm leak}(t).
\end{equation*}
For $F=0.1~\mu{\rm eV}/{\rm nm}$, the leakage reaches $P_{\rm leak}\sim 0.05$ on the time scale shown in Fig.~\ref{FigDynF0p1}(a), so that approximately $95\%$ of the total probability remains in the two discrete states. The populations $P_1$ and $P_2$ nevertheless retain a well-defined oscillatory transfer with $f_R=40~{\rm GHz}$. Thus, this case still exhibits predominantly two-level dynamics, although a $5\%$ leakage is already appreciable and should not be regarded as an optimal regime for high-fidelity manipulation. We therefore use $F=0.1~\mu{\rm eV}/{\rm nm}$ as an example close to the upper useful driving range. As shown below, reducing the amplitude to $F=0.05~\mu{\rm eV}/{\rm nm}$ suppresses the leakage below the percent level while preserving a Rabi frequency in the tens-of-GHz range.

\begin{figure*}
\includegraphics[width=1.85\columnwidth]{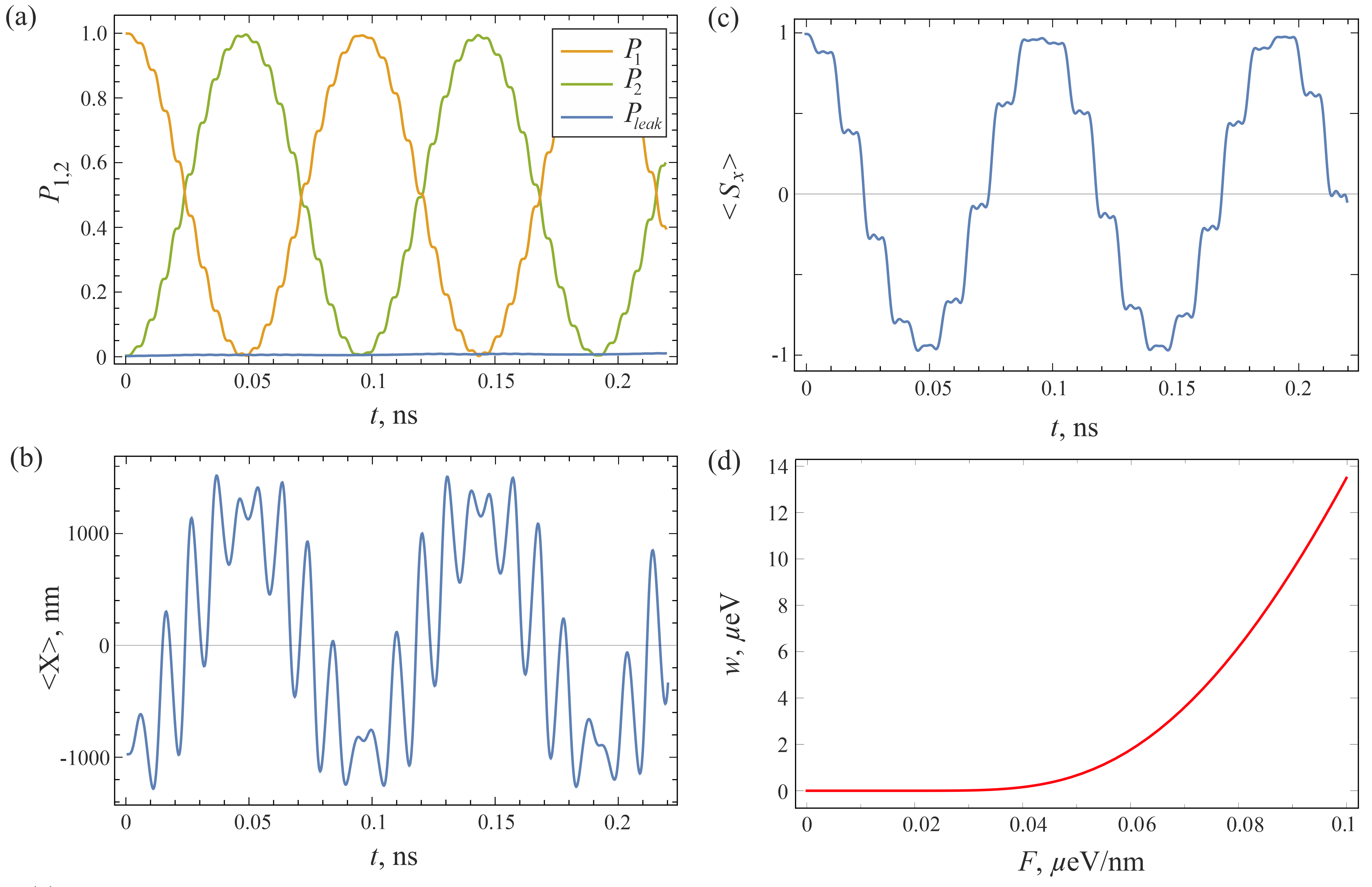}
\caption{\label{FigDynF0p05} Dynamics for the field amplitude $F=0.05~\mu$eV/nm, analogous to Fig.~\ref{FigDynF0p1}. (a) Populations $P_{1,2}(t)=|C_{1,2}(t)|^2$ and leakage $P_{\rm leak}=1-|C_1|^2-|C_2|^2$. The linear frequency of Rabi oscillations is $f_R=20$~GHz, and the leakage remains below $1\%$. (b,c) Time evolution of the expectation values $\langle X(t)\rangle$ and $\langle S_x(t)\rangle$. (d) Estimate~(\ref{wion}) for the leakage rate into the propagating 1D edge states as a function of $F$. When the field amplitude is reduced from $0.1$ to $0.05~\mu$eV/nm, the leakage rate decreases by more than one order of magnitude.}
\end{figure*}

Let us now consider the expectation values of physical observables for the state, described by $\Psi(x,t)$ in Eq.~(\ref{psit}). Figure~\ref{FigDynF0p1}(b) shows the time dependence of the mean coordinate $\langle X(t)\rangle$. It oscillates with the same frequency as the populations in Fig.~\ref{FigDynF0p1}(a), reflecting transitions between states localized predominantly near the left and right edge kinks. Similar Rabi-frequency oscillations are observed in the expectation value of the spin projection $\langle S_x(t)\rangle$, shown in Fig.~\ref{FigDynF0p1}(c). The state of the two-level subsystem can therefore be characterized either by its spatial localization or by the $x$ component of spin. The other spin components, $\langle S_y(t)\rangle$ and $\langle S_z(t)\rangle$, also oscillate but contain faster components, which can be suppressed by time averaging of the signal.

The spatially distinct probability distributions of the two localized states also suggest a possible experimental route for detecting the Rabi dynamics. As in conventional pulsed measurements of two-level systems, the coherent evolution does not need to be monitored continuously at the Rabi frequency. Instead, the system may be subjected to a resonant microwave pulse of variable duration $\tau$, followed by a transport readout of the final state. Repeating this sequence as a function of $\tau$ would provide access to the oscillating populations $P_1(\tau)$ and $P_2(\tau)$. In particular, a multiprobe transport configuration may be used to separate the contribution of the edge channel from the residual bulk conductivity~\cite{Benlemqwanssa2024}. We stress, however, that the quantitative relation between the occupations of the two localized states and a particular measured four-probe resistance depends on the detailed device and contact geometry and requires a separate transport analysis.

When the field amplitude is reduced from $0.1$ to $0.05~\mu$eV/nm, the leakage rate decreases by more than one order of magnitude. The effect of reducing the field amplitude is shown in Fig.~\ref{FigDynF0p05}, where the same quantities are plotted for $F=0.05~\mu$eV/nm. As seen from Fig.~\ref{FigDynF0p05}(a), the Rabi frequency is reduced approximately by a factor of two and becomes $f_R=20$~GHz. The same characteristic frequency appears in the oscillations of $\langle X(t)\rangle$ and $\langle S_x(t)\rangle$ in panels~(b) and (c). At the same time, the leakage into the propagating 1D states, shown by the lower curve in Fig.~\ref{FigDynF0p05}(a), is substantially smaller than for $F=0.1~\mu$eV/nm and remains below one percent over the considered time interval.

This strong suppression of leakage is related to the exponential dependence of the ionization rate $w(F)$ on the electric-field amplitude. In the simplest case of a potential well with a single discrete level, the ionization rate in energy units can be estimated as~\cite{wion}
\begin{equation}
w(F) \approx  2|E_{1}| \left( \frac{3F}{\pi F_0}\right)^{1/2} \exp \left(-\frac{2F_0}{3F}\right),
\label{wion}
\end{equation}
where $|E_1|$ is the absolute value of the discrete-level energy from which ionization occurs and $F_0$ characterizes the effective barrier field. In the present system it can be estimated as
\begin{equation}
F_0=\frac{\Delta_B^{(1)}-|E_1|}{L},
\label{f0}
\end{equation}
where $\Delta_B^{(1)}$ is given by Eq.~\eqref{deltab1}.

The estimate~(\ref{f0}) gives $F_0\sim 0.4~\mu$eV/nm. The resulting dependence~(\ref{wion}) is shown in Fig.~\ref{FigDynF0p05}(d). For $F=0.1~\mu$eV/nm, it yields an ionization rate of order
$w\sim 12~\mu$eV, corresponding to the leakage time $\tau\sim\hbar/w\sim 0.05$~ns. This estimate is consistent with the appearance of appreciable leakage in Fig.~\ref{FigDynF0p1}(a). When
the field is reduced to $F=0.05~\mu$eV/nm, the leakage rate decreases to approximately $w\sim 0.8~\mu$eV, and the corresponding leakage time becomes much longer than the characteristic Rabi period. As a result, the leakage is barely visible in Fig.~\ref{FigDynF0p05}(a).

The comparison of Figs.~\ref{FigDynF0p1} and~\ref{FigDynF0p05} therefore identifies the physically relevant operating regime of the double-kink system. The propagating 1D edge states do not preclude coherent two-level dynamics, but they impose an upper bound on the useful driving amplitude. Within this bound, the Rabi frequency remains in the tens-of-GHz range, while the leakage probability can be reduced to a level that is small on the time scale of several Rabi periods. Thus, the double-kink geometry realizes a lithographically defined two-level subsystem for studying electrically driven two-level dynamics together with coupling to extended edge states.

\section{\label{Sec:Conclusions} Results and conclusions}
We have studied localized and continuum states at the 1D edge of a HgTe/CdHgTe QW with two lithographically defined kinks. In a static in-plane magnetic field, the helical edge spectrum acquires a gap, and changes in the edge orientation modify the magnetic-field-induced edge mass. We have shown that, for suitable geometric parameters and magnetic-field direction, this produces two localized in-gap states at the kinks. These states form a lithographically defined two-level subsystem whose parameters are controlled by the edge shape, the kink separation, and the orientation of the magnetic field.

The main result of the work is that the localized two-level subsystem exhibits coherent electrically driven Rabi oscillations when coupling to the fully propagating 1D edge states  is included explicitly. For the parameters considered, the linear Rabi frequency reaches $20$--$40$~GHz. Population transfer into this continuum sector increases with the driving amplitude, whereas weaker fields strongly suppress the calculated leakage while preserving GHz-scale coherent oscillations. This trade-off defines an operating regime in which fast electric manipulation remains compatible with a well-preserved two-level dynamics.

The present analysis establishes a direct link between magnetic-field-induced corner-state physics and driven coherent dynamics of localized edge-state excitations. It also shows that the double-kink geometry provides more than a static pair of in-gap levels: it offers a controlled setting in which resonant transfer, electric-dipole coupling, and continuum-induced leakage can be analyzed within a single effective edge-state framework. The spatial separation of the two localized states further suggests a pulsed transport protocol in which resonant microwave manipulation is followed by a state-dependent edge-transport readout.

\begin{acknowledgments}
D.V.K and D.P.K. are supported by the State Assignment of the Ministry of Science and Higher Education of the Russian Federation, project No.~FSWR-2026-0004. S.S.K. and F.T. acknowledge financial support from the French Agence Nationale pour la Recherche through the ``Cantor'' (Grant No.~ANR-23-CE24-0022) and ``Teaser'' (Grant No.~ANR-24-CE24-4830) projects, as well as the Deutsche Forschungsgemeinschaft (DFG, German Research Foundation) through the W\"urzburg-Dresden Cluster of Excellence Complexity, Topology and Dynamics in Quantum Matter -- ctd.qmat (EXC 2147, project-id 390858490).
\end{acknowledgments}


%

\end{document}